\documentclass[aip,reprint]{revtex4-1}

\draft 
\usepackage{graphicx}
\usepackage[colorlinks=true,linkcolor=blue,citecolor=blue,urlcolor=blue]{hyperref}

\begin{document}


\title{Spin-squeezed states for metrology} 



\author{Alice Sinatra}
\email[]{alice.sinatra@lkb.ens.fr}
\homepage[]{http://www.phys.ens.fr/~sinatra/}
\affiliation{Laboratoire Kastler Brossel, ENS-Universit\'e PSL, CNRS, Universit\'e de la Sorbonne et Coll\`ege de France,
24 rue Lhomond, 75231 Paris, France}

\date{\today}

\begin{abstract}
Spin-squeezing is a well-established ``quantum technology'', where well-designed correlations in an ensemble of two-level systems reduce the statistical uncertainty of spectroscopic experiments. This paper reviews some important advances in the field, with emphasis on the author's contributions concerning in particular the fundamental limits imposed by decoherence. Building on the material presented in the first part, new ideas and some promising developments are outlined in the last section.
\end{abstract}


\maketitle 


\section{Introduction and overview}
\subsection{Collective spin, spin squeezing and quantum gain}
A quantum system with two energy levels $E_a$ and $E_b$ is formally equivalent to a spin $1/2$. Starting from $N$ such identical systems, for example atoms, one can form a collective spin operator $\vec{S}$ whose three components in first quantization are
\begin{equation}
\hat{S}_x+i\hat{S}_y={\small{ \sum_{i=1}^N}} | a\rangle \langle b |_i \:;\:
\hat{S}_z=\frac{\sum_{i=1}^N (|a\rangle \langle a|_i-|b\rangle \langle b|_i)}{2}
\end{equation}

In a state of the system in which each atom is in a superposition of the two levels, the collective spin precesses around the $z$-axis at the frequency $\omega_{ab}=(E_b-E_a)/\hbar$. Whether one wants to measure the unperturbed frequency $\omega_{ab}$ of a given atomic transition as in an atomic clock, or whether one is interested in the change of this frequency induced by an external field, for example a magnetic field, spin precession is at the heart of many atomic sensors.
\begin{figure}
\includegraphics[width=0.45\textwidth]{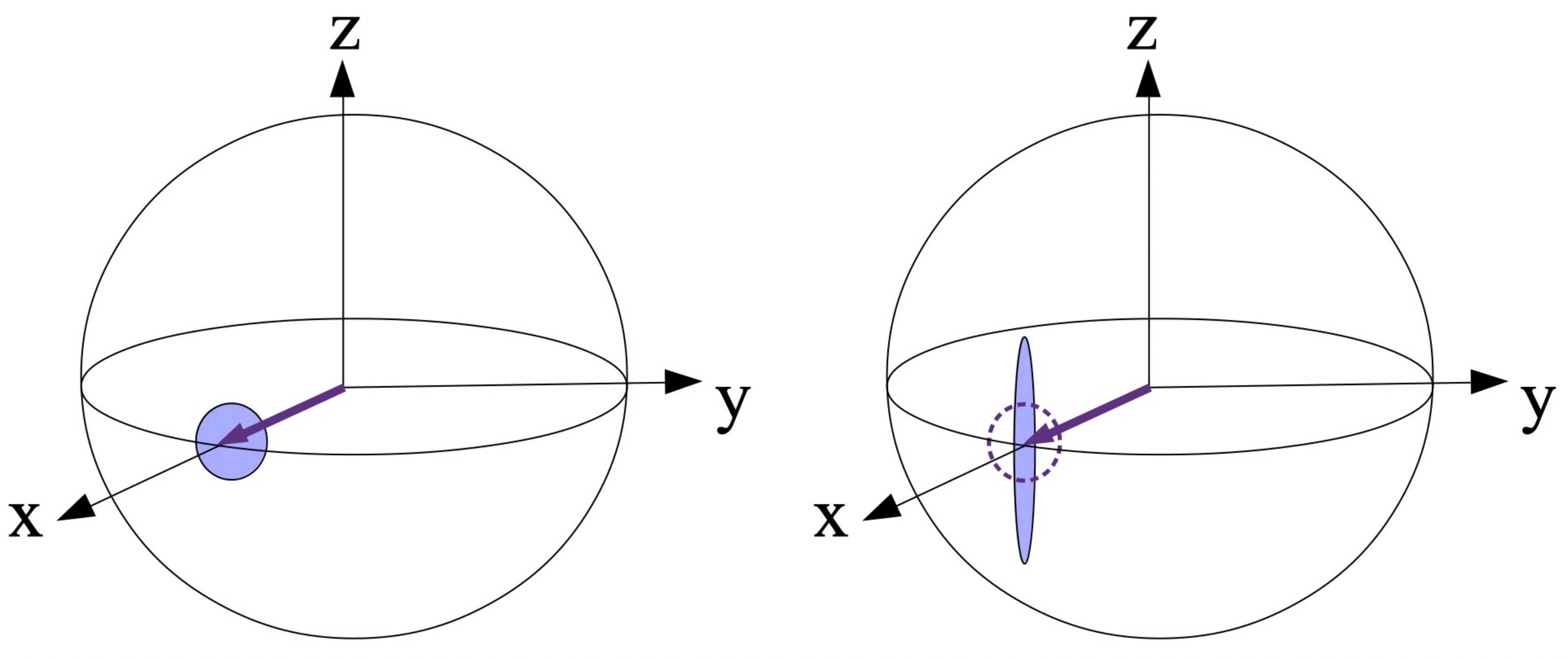}
\caption{
Collective spin prepared in the $\hat{S}_x$ eigenstate of eigenvalue $N/2$ without correlation between the atoms (left) and in a spin-squeezed state (right), represented on the Bloch sphere of radius $N/2$. The blue-colored part schematically represents the uncertainty on the spin components orthogonal to the mean spin, due to their quantum nature.
\label{fig:spin}}
\end{figure}
Such sensors, even after eliminating all sources of technical noise, remain limited by the quantum noise intrinsic to the state in which the collective spin has been prepared.  If, for example, the system is in the eigenstate of $\hat{S}_x$ with maximum eigenvalue $N/2$, tensor product of the state $(|a\rangle+|b\rangle)/\sqrt{2}$ for each atom shown on the left in figure \ref{fig:spin}, the uncertainty on the angular position of the collective spin is given by
\begin{equation}
(\Delta \phi)_{NC} = \frac{\Delta S_y}{\langle S_x \rangle} = \frac{\sqrt{N}}{N}=\frac{1}{\sqrt{N}} \,.
\end{equation}
On the contrary, in a spin-squeezed state \cite{KitagawaPRA1993}, shown on the right in figure \ref{fig:spin}, the uncertainty on one spin component orthogonally to the mean spin is reduced thanks to quantum correlations among the atoms. If the reduced uncertainty is on the $S_y$ component as in figure \ref{fig:spin}, the accuracy $\Delta \phi$ on the angular position of the collective spin is increased. We then define a spin squeezing parameter $\xi^2$,
\begin{equation}
\xi^2 \equiv \frac{N\Delta S_\perp^2}{|\langle \vec{S} \rangle|^2}
\label{eq:xi2}
\end{equation}
noise-to-signal squared ratio in a spectroscopy experiment where the phase $\phi$ is to be determined, which we want to be as small as possible, and which allows to quantify the metrological gain of a spin-squeezed state with respect to an uncorrelated one. In fact, we have \cite{WinelandPRA1992}
\begin{equation}
\Delta \phi=\xi (\Delta \phi)_{\rm NC} =\frac{\xi}{\sqrt{N}}
\label{eq:phi}
\end{equation}
The definition of $\xi^2$ in (\ref{eq:xi2}) can be generalized for non-Gaussian states \cite{FootnoteGauss} more complex than the spin-squeezed state represented in figure \ref{fig:spin}, for which the information on the precession phase $\phi$ must be extracted from an observable $\hat{X}$ which is in general a nonlinear function of the three components of the collective spin \cite{GessnerPRL2019,Baamara2021,Baamara2022}, with the same meaning as indicated by equation (\ref{eq:phi}).
\subsection{Non-linear evolution and non-destructive quantum measurement}
In order to generate the correlations that change a factorized state into a spin-squeezed state, an elegant method consists of a unitary evolution with a nonlinear Hamiltonian, e.g.\cite{KitagawaPRA1993}
\begin{equation}
H_{\rm NL}=\hbar \chi \hat{S}_z^2
\label{eq:OAT}
\end{equation}
which represents a Kerr-type nonlinearity for the atoms. It can result from short-range interactions between cold atoms in a Bose-Einstein condensate \cite{SorensenNATURE2001,EPJD}, but also, in an ensemble of atoms without interaction, from their collective interaction with an electromagnetic field in an optical cavity \cite{VuleticPRA2010}. In both cases it is possible to control the value of the constant $\chi$. For condensates, in the simple case where the states $|a\rangle$ and $|b\rangle$ play symmetric roles and atoms in different states do not interact with each other, $\chi=(d\mu/dN)/\hbar$ where $\mu=\rho g$ is the chemical potential and $\rho$ the atomic density for each of the two components. The coupling constant $g=4\pi\hbar^2a_s/m$ is proportional to the $s$-wave scattering length $a_s$ between cold atoms.

Another method to create a spin-squeezed state from an uncorrelated state consists in performing a quantum non-demolition measurement on a component of the collective spin. In general, this is done by coupling the system $\mathcal{S}$ to be measured to a second quantum system $\mathcal{P}$ and, after some interaction time between the two systems, by performing a destructive measurement on $\mathcal{P}$.
In order to measure non-destructively the $\hat{S}_z$ component of the collective atomic spin, one can take as a secondary system the Stokes spin $\hat{\vec{P}}$ of the light \cite{FootnoteStokes}, and use the Faraday interaction between the two spins \cite{KuzmichEPL,FootnoteBackaction}
\begin{equation}
H_{\rm Faraday}=\chi \hat{S}_z \hat{P}_z\,.
\label{eq:Fara}
\end{equation}

\subsection{Overview of the experimental results and application perspectives}
Using the two methods mentioned in the previous section, spin-squeezed states have been obtained in the laboratory and the demonstration of the their metrological gain has been done in several systems \cite{PezzeRMP2018}, which include Bose-Einstein gas condensates \cite{GrossNATURE2010, RiedelNATURE2010}, cold atoms interacting with an electromagnetic field \cite{LerouxPRL2010,MitchellPRL2012,HostenNATURE2016} for which a metrological gain $\xi^{-1}$ of the order of $10$ has been measured with $5\times10^5$ atoms \cite{HostenNATURE2016}, and hot atomic vapors \cite{BaoScience2020}. These remarkable results are proofs of principle. The transition to applications in metrology should take place if quantum gain becomes easily accessible from a technological point of view or if in some cases it is indispensable. Among the various atomic sensors, the clocks that already in 1999 had reached and observed the quantum limit in the case of independent atoms \cite{SantarelliPRL1999}, could be the first to benefit from spin squeezing techniques. In general, the smaller the number $N$ of atoms that we use, and here we have in mind the evolution towards miniaturisation of atomic sensors, the greater the relative importance of quantum noise and the potential utility of spin-squeezed states.

\section{Quantum Metrology and Decoherence}
Quantum metrology aims to improve the accuracy of measurements by reducing their ``fundamental'' statistical uncertainty given by quantum fluctuations. This is achieved by preparing the physical system of interest in a state with particular quantum correlations. For this approach to make sense from a practical point of view, two conditions seem necessary.
The first is that the statistical uncertainty on the variable, obtained from a measurement with independent atoms in an imparted time, is dominated by quantum fluctuations and not by other sources of classical noise. The second is that the decoherence is sufficiently weak not to destroy the quantum correlations we want to put to use. Mathematically, if we fix the time $T$ of the measurement, the second condition is automatically verified in the limit $\gamma \to 0$ where $\gamma$ is the decoherence rate in the system.

\subsection{Scaling laws of the quantum gain}
If, as we have said, decoherence plays an important role in the use of entangled states, it equally plays an important role in their preparation, determining in general a fundamental limit to the quantum gain that can be obtained.
Starting from an uncorrelated, eigenstate of $\hat{S}_x$ of eigenvalue $N/2$ with $N\gg1$, unitary evolution with the nonlinear Hamiltonian (\ref{eq:OAT}) generates at an optimal time $\chi t \simeq N^{-2/3}$ a spin-squeezed state for which $\xi^2 \simeq N^{-2/3}$.
This very favourable scaling law of quantum gain for spin-squeezed states becomes even more favorable for non-Gaussian states obtained by prolonging the unitary evolution for longer times until the maximum limit $\xi^2=N^{-1}$ is reached. \cite{ChalopinNCom2018,SchleierSmithPRL2016,VuleticArxiv2021} However, very entangled states offering large quantum gain are in general more fragile to decoherence, which modifies the scaling laws of quantum gain \cite{Kittens,Baamara2021,Baamara2022}. In some cases one loses the scaling law completely and obtains a spin squeezing parameter that tends towards a constant $\xi^2_{\rm min}$ in the limit $N\to\infty$. If this indicates that one finds a classical scaling law for $\Delta \phi$ in the light of the equation $\ref{eq:phi}$, the quantum gain can remain very large if $\xi^2_{\rm min}\ll1$.
\section{Fundamental limits for spin squeezing in Bose-Einstein condensates}

With colleagues we have studied the fundamental limits on spin squeezing that can be achieved in a Bose-condensed gas formed by $N$ two-level atoms, usually two hyperfine levels in the fundamental state, using short-range van der Waals interactions between cold atoms. While this physical system accurately realizes the simple model (\ref{eq:OAT}) in the mean-field description \cite{SorensenNATURE2001,EPJD}, an intrinsic source of decoherence for spin squeezing dynamics comes from the interaction of the condensate with thermally populated low-energy excitations in the gas, whose population fluctuates from one realization to another in the experiment even under identical conditions for the experimental parameters. In the framework of the multimodal microscopic description of Bogoliubov \cite{Bogoliubov}, we have shown that at the thermodynamic limit $N\to \infty$, and for an optimal interaction time $t_{\rm opt}$ which remains finite, the squeezing parameter tends towards a constant $\xi^2_{\rm min}$:  \cite{SinatraPRL2011}
\begin{equation}
\xi^2(t_{\rm opt}) \stackrel{N\to\infty}{\to} \xi^2_{\rm min} \,,
\label{eq:xiopt}
\end{equation}
which then constitutes a limit for the quantum gain.
For a temperature above the chemical potential of the gas, but well below the critical condensation temperature, the limit $\xi^2_{\rm min}$ is well approximated by the initial non-condensed fraction in the gas. An important point is that for realistic values of the parameters, it is in principle possible to obtain values of $\xi_{\rm min}^2$ of the order of $1/900$ corresponding to a gain of a factor $30$ on $\Delta \phi$ according to the formula (\ref{eq:phi}) compared to the use of uncorrelated atoms.

A second source of fundamental decoherence that we have considered is given by particle loss, a consequence of an imperfect vacuum for one-body losses, and inherent to the metastable nature of ultracold gases for three-body losses. In the theoretical framework of an open system described by a master equation \cite{LiYunPRL2008,SinatraFro2012} we have shown that in the limit of $N$ large the optimal spin squeezing is limited by a constant, as in (\ref{eq:xiopt}), where
$\xi^2_{\rm min}$ is in this case proportional to the fraction of atoms lost at time $t_{\rm opt}$.
Once spin squeezing has been prepared at time $t=0$, its evolution, in the absence of interactions and in the presence of one-body losses of rate $\gamma$, is given by
\begin{equation}
\xi^2(t)-1=\left[\xi^2(0)-1 \right]e^{-\gamma t} \,.
\label{eq:evloss}
\end{equation}
Spin squeezing that lasts more than one second and is limited by losses according to equation (\ref{eq:evloss}) has been seen in \cite{MylesArXiv}.

\section{Promising developments for the future}
Research in the field of quantum technologies is abundant, and several promising avenues are currently being explored. For atomic clocks, two significant examples are the first demonstration of spin-squeezing on an optical transition \cite{VuleticNat2020} and the theoretical studies of spin-squeezing with atoms in optical lattices \cite{EPLSqMott,AMRlattice}. The purpose of this section is to outline some promising developments from the material we presented in the first part of the article. These ideas fit perfectly into the theoretical framework described so far, as they are based on either non-linear evolution (\ref{eq:OAT}) or quantum non-demolition Faraday interaction (\ref{eq:Fara}). However, due to the difference in the physical system used, the time scales involved or the type of quantum platform that can be obtained, the realization of each of them would represent a breakthrough from the state of the art and open the field to new applications.

\begin{figure}
\includegraphics[width=0.4\textwidth]{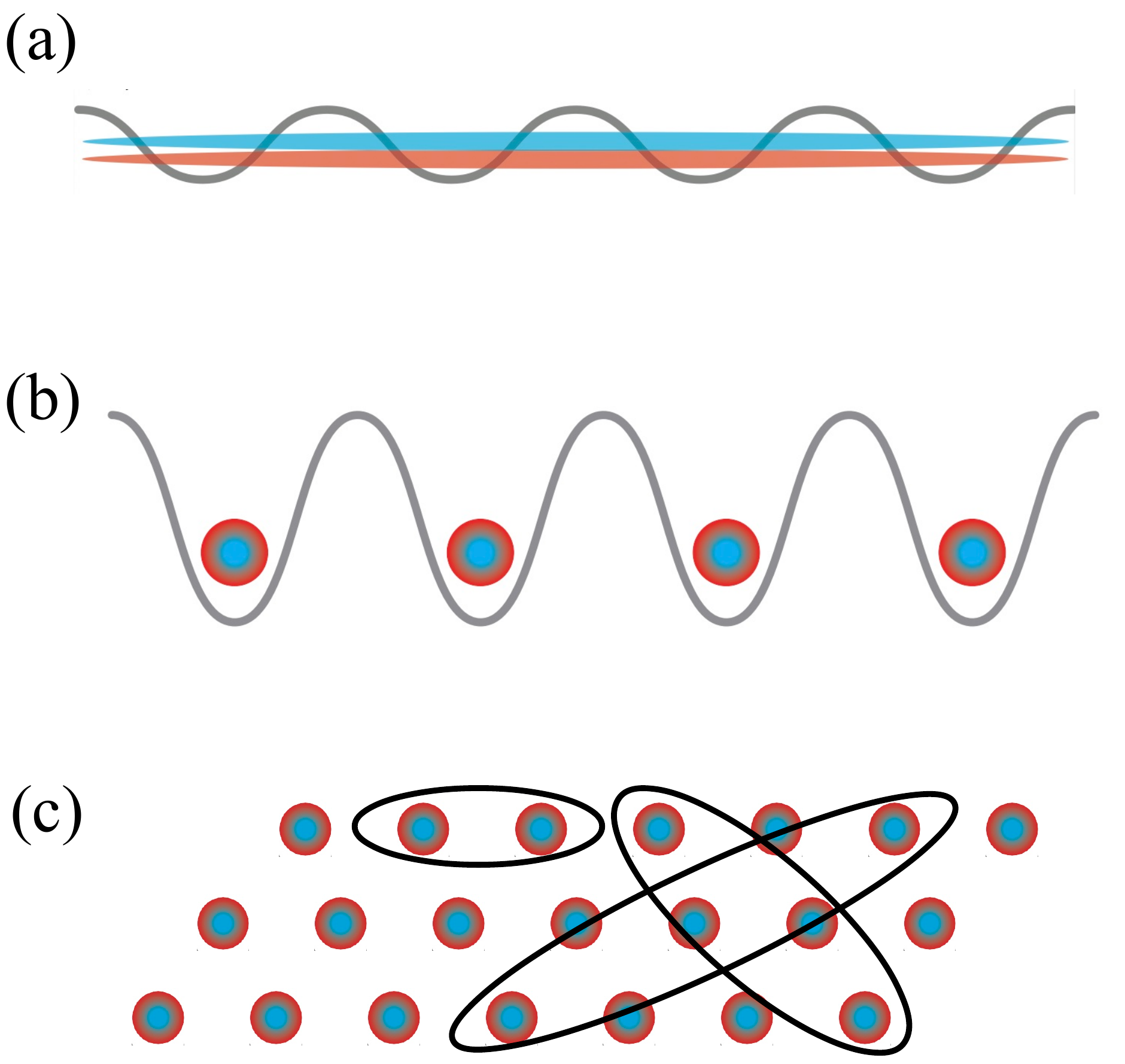}
\caption{(a) An optical lattice is adiabatically raised in an interacting two-component condensate. (b) The system enters the Mott-insulating phase with one atom per site at the optimal squeezing time, thus freezing the quantum correlations between atoms (figure adapted from \cite{EPLSqMott}). (c) In the squeezed-Mott configuration, each atom is in a superposition of the two the two internal states $|a\rangle$ and $|b\rangle$ and is equally correlated to any other atom. The entanglement is then distributed between spatially separated and individually accessible modes.}
\label{fig:SqMott}
\end{figure}

{\it Storing spin-squeezing in a Mott-insulator phase -} We have already mentioned that the binary short-ranged van der Waals interactions in a Bose-Einstein condensate where each atom is prepared in a superposition of two internal states $(|a\rangle+|b\rangle)/\sqrt{2}$, generate the nonlinear Hamiltonian (\ref{eq:OAT}) that transforms the factorized initial state into a spin-squeezed state. Physically, since the interaction strength depends on the atoms internal state, the {\it phase of each atom}, represented by the $\hat{S}_y$ component of the collective spin, correlates {\it via} the mean field shift with the {\it population difference} of the ensemble, represented by $\hat{S}_z$. For phase sensing or atom interferometry, the squeezing, that is hence generated in a superposition of $\hat{S}_y$ and $\hat{S}_z$, can be transferred to the $\hat{S}_y$ spin component, as in figure \ref{fig:spin}, by a simple rotation. 
To this physical picture we now add an optical lattice that is adiabatically raised in the gas, progressively bringing the interacting system from the superfluid to the Mott-insulator phase \cite{Zoller1998}. As long as there is a delocalised superfluid component, figure \ref{fig:SqMott} (a), the spin-squeezing dynamics continues to create correlations among the internal states of the atoms. When the system enters the Mott-insulator phase with one atom per site, figure \ref{fig:SqMott} (b), the condensate disappears, the van der Waals interactions are suppressed as the atoms are separated, but the correlations among the internal states of the atoms that were created in the superfluid phase remain. We have shown in this way that one can freeze a spin-squeezed state \cite{EPLSqMott} or an even more entangled non-Gaussian state \cite{PRASqMott}, in a Mott-insulator configuration where each atom is a spin $1/2$ that shares quantum correlations with all others, figure \ref{fig:SqMott} (c). 
 
In addition to the potential interest for atomic clocks \cite{KatoriNP,JYeScience2017}, such a configuration where quantum entanglement is shared between individually accessible spatial modes offers interesting prospects for quantum information and multiparameter quantum metrology \cite{MultiManuel1,MultiManuel2}.

\begin{figure}
\includegraphics[width=0.5\textwidth]{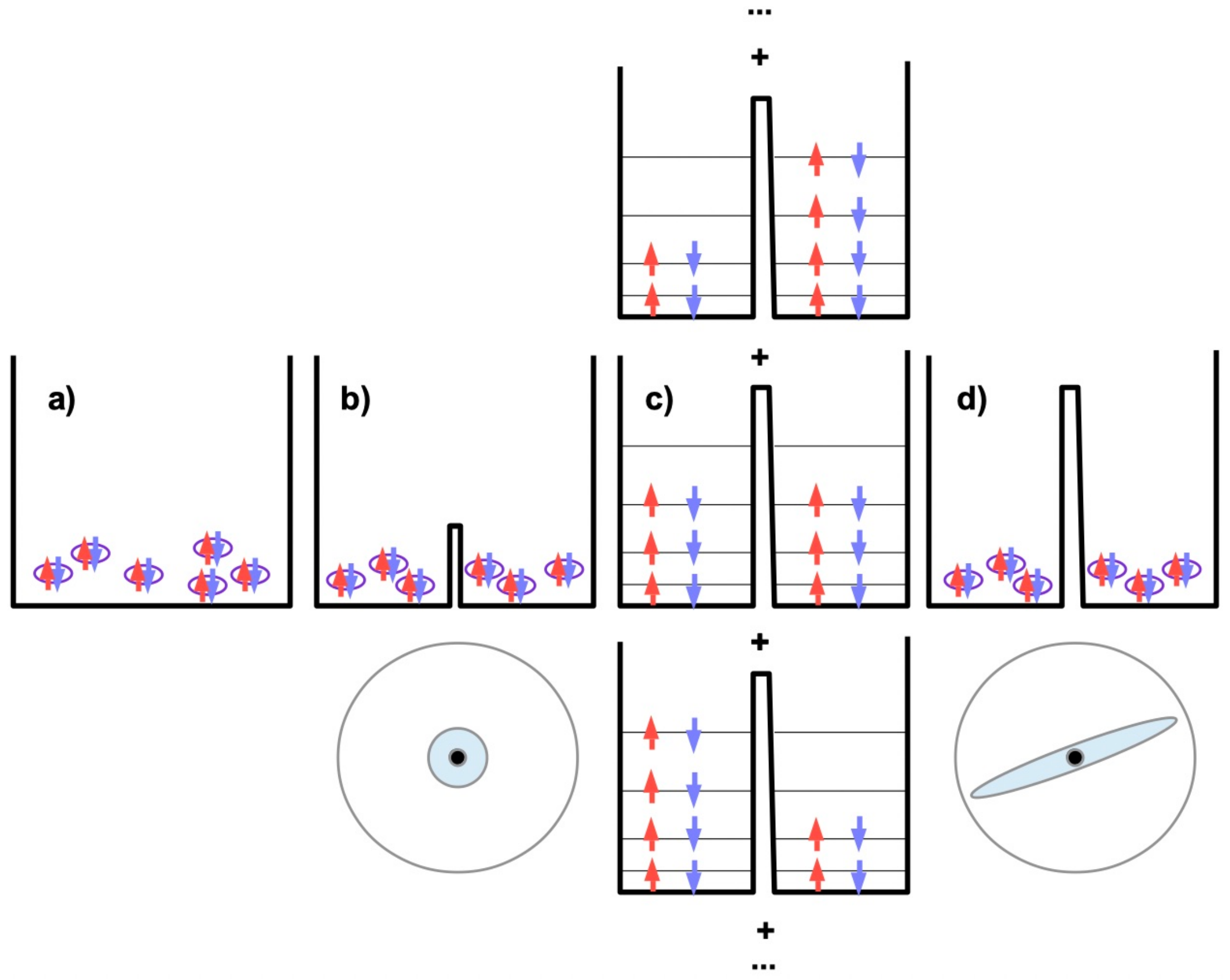}
\caption{Spin squeezing Gedankenexperiment with a condensate of fermion pairs at zero temperature in a double-well. (a)-(b) A condensate of fermions ``dimers" is prepared in a delocalized coherent state on the two wells. (c) After raising adiabatically the barrier separating the two wells, the scattering length is set to the desired value, e.g. $a_s \to 0^-$, for the spin squeezing phase. In this case, squeezing is created in the ideal Fermi gas due to the $N$-dependence of the chemical potential: $\epsilon_F \propto \rho^{2/3}$ in 3D. After the squeezing phase, the scattering length is restored to the initial positive value corresponding to a dimer condensate, obtaining a spin-squeezed state for the latter. In the lower parts of figures (b) and (d), we plot on the Bloch sphere the fluctuations of the collective spin constructed from the $a$ and $b$ modes for the center of mass of the dimers corresponding to the ground state in the left and right wells.}
\label{fig:Fermi}
\end{figure}

{\it Using the non linearity of a degenrate Fermi gas -} Interesting perspectives are also offered by degenerate gases of fermions in two spin states, in which one can control the strength of the interactions without inducing significant losses of atoms, and obtain at will: a gas of dimers formed by two strongly bonded atoms of opposite spin, a gas of weakly bonded Cooper pairs, or a strongly interacting gas in the so-called unitary regime \cite{ZwergerBook}.             
To give an example, let us consider the following {\it Gedankenexperiment} inspired by our studies on phase coherence of a paired-fermions condensate \cite{PRAphase,CRphase}. A condensate of fermion dimers of opposite spin is initially prepared in a symmetric double-well in an uncorrelated state, where each dimer is symmetrically delocalized on the two wells (figure \ref{fig:Fermi}(a)-(b)). By raising the barrier between the two wells until they are completely separated, slowly enough not to excite the condensates, one actually prepares the initial state $\langle \hat{S}_x \rangle=N/2$ where this time $a$ and $b$ refer to the two potential wells, and one can show that the state of the system evolves according to the Hamiltonian (\ref{eq:OAT}) where, as for atomic condensates, $\chi=(d\mu/dN)/\hbar$, $\mu$ being the chemical potential of the gas in each well.
Note that here $\hat{S}_z$ represents the difference between the number of pairs in the two wells, $\hat{S}_x$ represents the macroscopic phase coherence between the two wells and, again, the quantity that is squeezed by the Hamiltonian (\ref{eq:OAT}) is a superposition of $\hat{S}_z$ and $\hat{S}_y$.
Similar experiments with bosonic condensates, aimed at using spin-squeezed states in an atomic interferometer, have been carried out \cite{Obethaler1,Berrada2013}. In the case of the fermionic-pairs condensate, as shown for example in \cite{PRAphase}, the largest value of $\chi$, hence the fastest spin squeezing dynamics, is obtained in the limit $a_s\to 0^-$ where the chemical potential in each well tends towards the Fermi energy of an ideal gas of fermions $\mu \to \epsilon_F$. This paradoxical situation, where spin squeezing would be generated in a Fermi sea without interaction, is depicted in figure \ref{fig:Fermi}(c).
In the homogeneous three-dimensional case, it is straightforward to compare the nonlinearities $\chi_{\rm Bose}=\frac{\rho g}{N}$ and $\chi_{\rm Fermi}=\frac{2}{3}\frac{\epsilon_F}{N}$ with $\epsilon_F\propto \rho^{2/3}$ and it is seen that the nonlinearity is much larger for the ideal Fermi gas
\begin{equation}
\frac{\chi^{\rm Fermi}}{\chi^{\rm Bose}}\simeq 
\frac{\epsilon_F}{\rho g}\simeq 
\frac{1}{a_s\rho^{1/3}}\gg 1
\end{equation}
proportionally to the power $-\frac{2}{3}$ of the parameter $\sqrt{\rho a_s^3}$, which typically holds $10^{-3}$-$10^{-2}$ in an atomic Bose-Einstein condensate \cite{Lopez2017}. If these considerations are very encouraging, a study at non-zero temperature is needed to draw conclusions on the best interaction regime to use with fermions
and calculate the maximum spin squeezing that can be achieved with this method.

{\it Long-lived nuclear spin squeeing in a gas cell -} In the field of atomic sensors, spin squeezing and quantum technologies have a vocation to extend to room temperature gases or vapours in a cell \cite{MitchellNC2020,macQsimal}. Although the physical system is different from those discussed so far, once the average over the rapid movement of the atoms in the cell has been taken, it is possible to describe the internal degrees of freedom in terms of the collective spin and connect to the theoretical description presented in the article. 
An extreme system for its level of isolation and coherence time, measured in tens of hours, is the purely nuclear collective spin of a rare gas such as helium-3 in its ground state.
A first theoretical study showed that exchange collisions between helium atoms in the ground state and in an optically accessible metastable state, currently used to polarize the nuclear spins \cite{GentileNacher2017}, allow quantum correlations and squeezing to be transferred to the nuclear spin \cite{Dantan2005a,Reinaudi2007a}. More recently, it has been shown that spin squeezing schemes using a quantum non demolition measurement  with a Faraday Hamiltonian of type (\ref{eq:Fara}), successfully realized for alkaline gas \cite{stroboscopic}, can be adapted to nuclear spin squeezing \cite{FirstenbergPolzik,SerafinPRL2021,SerafinCR2021}, using spin-exchange collisions between the rare gas and an alkaline \cite{KornackRomalis,OferNatPhys} or metastability exchange collisions in a pure helium-3 gas. \cite{LaloeDupontLeduc} In the latter case, we have predicted that the best spin squeezing parameter $\xi^2$ accessible \cite{SerafinCR2021} scales as $\sqrt{\gamma_0/\gamma_m}$ where $\gamma_m$ is the rate of metastability exchange collisions for a metastable helium atom, of the order of $5\times10^6s^{-1}$ for a ground state helium pressure of $2$ mbar, and $\gamma_0$ is the decoherence rate of metastable atoms, dominated by their diffusion towards the cell walls, of the order of $2\times 10^4s^{-1}$ for a centimetric cell without buffer gas. The numerous spectroscopic applications of helium nuclear spin, ranging from magnetometry to fundamental physics tests \cite{GentileNacher2017}, could thus in the future benefit from quantum noise control and stimulate new ideas and applications of quantum technologies.

\section{Conclusion}
Spin squeezing and quantum metrology is a very active field, constantly fed by new ideas, extensive theoretical studies and surprising experimental discoveries. The transition from science to technology should take place when the quantum advantage outweighs the complexity of the experiments or, as we have said, if it is indispensable in some cases. An interesting example from this point of view is that of the squeezed states of light, extensively studied in the 1980s and now used in an interferometer to detect gravitational waves \cite{TsePRL2019}.

\begin{acknowledgments}
We thank Yvan Castin for his valuable comments on the manuscript. 
\end{acknowledgments}


\begin{thebibliography}{0}%
\makeatletter
\providecommand \@ifxundefined [1]{%
 \@ifx{#1\undefined}
}%
\providecommand \@ifnum [1]{%
 \ifnum #1\expandafter \@firstoftwo
 \else \expandafter \@secondoftwo
 \fi
}%
\providecommand \@ifx [1]{%
 \ifx #1\expandafter \@firstoftwo
 \else \expandafter \@secondoftwo
 \fi
}%
\providecommand \natexlab [1]{#1}%
\providecommand \enquote  [1]{``#1''}%
\providecommand \bibnamefont  [1]{#1}%
\providecommand \bibfnamefont [1]{#1}%
\providecommand \citenamefont [1]{#1}%
\providecommand \href@noop [0]{\@secondoftwo}%
\providecommand \href [0]{\begingroup \@sanitize@url \@href}%
\providecommand \@href[1]{\@@startlink{#1}\@@href}%
\providecommand \@@href[1]{\endgroup#1\@@endlink}%
\providecommand \@sanitize@url [0]{\catcode `\\12\catcode `\$12\catcode
  `\&12\catcode `\#12\catcode `\^12\catcode `\_12\catcode `\%12\relax}%
\providecommand \@@startlink[1]{}%
\providecommand \@@endlink[0]{}%
\providecommand \url  [0]{\begingroup\@sanitize@url \@url }%
\providecommand \@url [1]{\endgroup\@href {#1}{\urlprefix }}%
\providecommand \urlprefix  [0]{URL }%
\providecommand \Eprint [0]{\href }%
\providecommand \doibase [0]{http://dx.doi.org/}%
\providecommand \selectlanguage [0]{\@gobble}%
\providecommand \bibinfo  [0]{\@secondoftwo}%
\providecommand \bibfield  [0]{\@secondoftwo}%
\providecommand \translation [1]{[#1]}%
\providecommand \BibitemOpen [0]{}%
\providecommand \bibitemStop [0]{}%
\providecommand \bibitemNoStop [0]{.\EOS\space}%
\providecommand \EOS [0]{\spacefactor3000\relax}%
\providecommand \BibitemShut  [1]{\csname bibitem#1\endcsname}%
\let\auto@bib@innerbib\@empty
\end{thebibliography}%


%


\begin{thebibliography}{99}

\bibitem{KitagawaPRA1993} M. Kitagawa and M. Ueda, Squeezed spin states, \href{https://doi.org/10.1103/PhysRevA.47.5138}{Phys. Rev. A. \textbf{47}, 5138 (1993)}.

\bibitem{WinelandPRA1992} D. J. Wineland, J. J. Bollinger, W. M. Itano, F. L. Moore, and
D. J. Heinzen, Spin squeezing and reduced quantum noise in spectroscopy, \href{https://doi.org/10.1103/PhysRevA.46.R6797}{Phys. Rev. A \textbf{46}, R6797 (1992)}.

\bibitem{FootnoteGauss} By non-Gaussian states, we mean spin states for which the Wigner representation on the Bloch sphere \cite{PezzeRMP2018} is non-Gaussian and has fine structures that make the state more sensitive to rotation.

\bibitem{GessnerPRL2019} M. Gessner, A. Smerzi and L. Pezz{\`e}, Metrological Nonlinear Squeezing Parameter, \href{https://doi.org/10.1103/PhysRevLett.122.090503}{Phys. Rev. Lett. \textbf{122}, 090503 (2019)}.

\bibitem{Baamara2021} Y. Baamara, A. Sinatra, and M. Gessner, Scaling Laws for the Sensitivity Enhancement of Non-Gaussian Spin States, \href{https://doi.org/10.1103/PhysRevLett.127.160501}{Phys. Rev. Lett. \textbf{127}, 160501 (2021)}.

\bibitem{Baamara2022} Y. Baamara, A. Sinatra, and M. Gessner, Squeezing of nonlinear spin observables by one axis twisting in the presence of decoherence: An analytical study, 
\href{https://comptes-rendus.academie-sciences.fr/physique}{Comptes Rendus Physique, {\it to appear}}.

\bibitem{SorensenNATURE2001} A. S\o{}rensen, L. M. Duan, J. I. Cirac, and P. Zoller, Many-particle entanglement with Bose-Einstein condensates, \href{http://dx.doi.org/10.1038/35051038}{Nature \textbf{409}, 63 (2001)}.

\bibitem{EPJD} Y. Li, P. Treutlein, J. Reichel and A. Sinatra, Spin squeezing in a bimodal condensate: spatial dynamics and particle losses, \href{https://doi.org/10.1140/epjb/e2008-00472-6}{Eur. Phys. J. B {\bf 68}, 365 (2009)}.

\bibitem{VuleticPRA2010} M. H. Schleier-Smith, I. D. Leroux and V. Vuleti{\`c}, Squeezing the collective spin of a dilute atomic ensemble by cavity feedback, \href{https://doi.org/10.1103/PhysRevA.81.021804}{Phys. Rev. A {\bf 81}, 021804 (2010)}.

\bibitem{FootnoteStokes} The Stokes spin $\hat{\vec{P}}$ is obtained from the two polarisation modes $x$ and $y$ of an electromagnetic field propagating in the $z$-direction $P_x= \frac{1}{2}\left( c_x^\dagger c_x -  c_y^\dagger c_y \right)$ and $P_y+iP_z = c_x^\dagger c_y$, where $c_\alpha$ is the annihilation operator of a photon polarized in the direction $\alpha=x,y$.

\bibitem{KuzmichEPL} A. Kuzmich, N. P. Bigelow and L. Mandel, Atomic quantum non-demolition measurements and squeezing,
\href{https://doi.org/10.1209/epl/i1998-00277-9}{Europhys. Lett. {\bf 42} 481(1998)}.

\bibitem{FootnoteBackaction} When the two spins are initially polarized along the $x$-direction, with $\hat{P}_x\simeq n_{\rm ph}/2$ where $n_{\rm ph}$ is the number of photons in the mode and $\hat{S}_x\simeq N/2$ where $N$ is the number of atoms, the $y$-component of the Stokes spin correlates with the $z$-component of the atomic spin since $[\hat{P}_y,H_{\rm Faraday}] \simeq \chi (n_{\rm ph}/2) \hat{S}_z$,  thus allowing the measurement  of $\hat{S}_z$, while the back-action of the measurement affects the $y$-component of the atomic spin since $[\hat{S}_y,H_{\rm Faraday}] \simeq \chi (N/2) \hat{P}_z$.

\bibitem{PezzeRMP2018} L. Pezz\`{e}, A. Smerzi, M. K. Oberthaler, R. Schmied, and P. Treutlein,
Quantum metrology with nonclassical states of atomic ensembles, \href{https://doi.org/10.1103/RevModPhys.90.035005}{Rev. Mod. Phys. \textbf{90}, 035005 (2018)}.

\bibitem{GrossNATURE2010} C. Gross, T. Zibold, E. Nicklas, J. Est{\`e}ve, and M. K. Oberthaler, Nonlinear atom interferometer surpasses classical precision limit, \href{https://doi.org/10.1038/nature08919}{Nature \textbf{464}, 1165 (2010)}.

\bibitem{RiedelNATURE2010} M. F. Riedel, P. Böhi, Y. Li, T. W. Hänsch, A. Sinatra, and P. Treutlein, Atom-chip-based generation of entanglement for quantum metrology, \href{https://doi.org/10.1038/nature08988}{Nature \textbf{464}, 1170 (2010)}.

\bibitem{LerouxPRL2010}
I. D. Leroux, M. H. Schleier-Smith, and V. Vuleti{\`c},
Implementation of cavity squeezing of a collective atomic spin,
\href{https://doi.org/10.1103/PhysRevLett.104.073602}{Phys. Rev. Lett. {\bf 104}, 073602 (2010)}.

\bibitem{MitchellPRL2012} R. J. Sewell, M. Koschorreck, M. Napolitano, B. Dubost, N. Behbood, and M. W. Mitchell, Magnetic Sensitivity Beyond the Projection Noise Limit by Spin Squeezing, \href{https://doi.org/10.1103/PhysRevLett.109.253605}{Phys. Rev. Lett. \textbf{109}, 253605 (2012)}.

\bibitem{HostenNATURE2016} O. Hosten, N. J. Engelsen, R. Krishnakumar and M. A. Kasevich, Measurement noise 100 times lower than the quantum-projection limit using entangled atoms, \href{https://doi.org/10.1038/nature16176}{Nature \textbf{529}, 505 (2016)};

\bibitem{BaoScience2020} 
H. Bao, J. Duan, S. Jin \textit{et al.}, Spin squeezing of 1011 atoms by prediction and retrodiction measurements, \href{https://doi.org/10.1038/s41586-020-2243-7}{Nature \textbf{581}, 159 (2020)}.

\bibitem{SantarelliPRL1999} G. Santarelli, Ph. Laurent, P. Lemonde, A. Clairon, A. G. Mann, S. Chang, A. N. Luiten, A. N. and C. Salomon, Quantum Projection Noise in an Atomic Fountain: A High Stability Cesium Frequency Standard, 
 \href{https://link.aps.org/doi/10.1103/PhysRevLett.82.4619}{Phys. Rev. Lett. \textbf{82}, 4619 (1999)}

\bibitem{SchleierSmithPRL2016} E. Davis, G. Bentsen and M. Schleier-Smith, Approaching the Heisenberg Limit without Single-Particle Detection, \href{https://doi.org/10.1103/PhysRevLett.116.053601}{Phys. Rev. Lett. \textbf{116}, 053601 (2016)}.

\bibitem{ChalopinNCom2018}
T. Chalopin, C. Bouazza, A. Evrard, V. Makhalov, D. Dreon, J. Dalibard, L. A. Sidorenkov, S. Nascimbene,
Quantum-enhanced sensing using non-classical spin states of a highly magnetic atom, \href{https://www.nature.com/articles/s41467-018-07433-1}{Nat. Commun. {\bf 9}, 4955 (2018).}

\bibitem{VuleticArxiv2021} S. Colombo, E. Pedrozo-Pe{\~n}afiel, A. F. Adiyatullin, Z. Li, E. Mendez, C. Shu and V. Vuleti{\`c}, Time-Reversal-Based Quantum Metrology with Many-Body Entangled States, \href{https://arxiv.org/abs/2106.03754}{arXiv:2106.03754}.

\bibitem{Kittens} K. Pawlowski, M. Fadel, P. Treutlein, Y. Castin, and A. Sinatra, Mesoscopic quantum superpositions in bimodal Bose-Einstein condensates: Decoherence and strategies to counteract it, \href{https://doi.org/10.1103/PhysRevA.95.063609}{Phys. Rev. A {\bf 95}, 063609 (2017)}.

\bibitem{Bogoliubov} N.N. Bogolioubov, On the theory of superfluidity, 
\href{https://ufn.ru/pdf/jphysussr/1947/11_1/3jphysussr19471101.pdf}{J. Phys. (USSR) {\bf 11}, 23 (1947)}.

\bibitem{SinatraPRL2011} A. Sinatra, E. Witkowska, J.-C. Dornstetter, Y. Li, and Y. Castin, Limit of Spin Squeezing in Finite-Temperature Bose-Einstein Condensates, \href{https://doi.org/10.1103/PhysRevLett.107.060404}{Phys. Rev. Lett. \textbf{107}, 060404 (2011)}.

\bibitem{LiYunPRL2008} Y. Li, Y. Castin, and A. Sinatra, Optimum Spin Squeezing in Bose-Einstein Condensates with Particle Losses, \href{https://doi.org/10.1103/PhysRevLett.100.210401}{Phys. Rev. Lett. \textbf{100}, 210401 (2008)}.

\bibitem{SinatraFro2012} A. Sinatra, J.-C. Dornstetter and Y. Castin, Spin squeezing in Bose-Einstein condensates: Limits imposed by decoherence and non-zero temperature, \href{https://doi.org/10.1007/s11467-011-0219-7}{Front. Phys. \textbf{7}, 86 (2012)}.

\bibitem{MylesArXiv} 
M.-Z. Huang, J.~A. de~la Paz, T.~Mazzoni, K.~Ott, A.~Sinatra, C.~L.~G. Alzar, J.~Reichel, 
Self-amplifying spin measurement in a long-lived spin-squeezed state, \href{https://arxiv.org/abs/2007.01964}{arXiv:2007.01964}.  

\bibitem{VuleticNat2020} E. P.-Pe\~nafiel, S. Colombo, C. Shu, A. F. Adiyatullin, Z. Li, E. Mendez, B. Braverman, A. Kawasaki, D. Akamatsu, Y. Xiao, and V. Vuleti{\'c}, Entanglement on an optical atomic-clock transition,  \href{https://doi.org/10.1038/s41586-020-3006-1}{Nature, \textbf{414}, 588 (2020).}

\bibitem{EPLSqMott} D. Kajtoch, E. Witkowska, A. Sinatra, Spin-squeezed atomic crystal
\href{https://doi.org/10.1209/0295-5075/123/20012}{Europhys. Lett {\bf 123}, 20012 (2018)}.

\bibitem{AMRlattice} P. He, M. A. Perlin, S. R. Muleady, R. J. Lewis-Swan,  R. B. Hutson, J. Ye, and A. M. Rey,
Engineering spin squeezing in a 3D optical lattice with interacting spin-orbit-coupled fermions, 
\href{https://doi.org/10.1103/PhysRevResearch.1.033075}{Phys. Rev. Res. {\bf 1}, 033075 (2019)}.

\bibitem{Zoller1998} D. Jaksch, C. Bruder, J. I. Cirac, C. W. Gardiner, and P. Zoller, 
Cold Bosonic Atoms in Optical Lattices,  \href{https://doi.org/10.1103/PhysRevLett.81.3108}{Phys. Rev. Lett. \textbf{81}, 3108 (1998)}.

\bibitem{PRASqMott} M. Plodzien, M. Koscielski, E. Witkowska, A. Sinatra, Producing and storing spin-squeezed states and Greenberger-Horne-Zeilinger states in a one-dimensional optical lattice, \href{https://doi.org/10.1103/PhysRevA.102.013328}{Phys. Rev. A 102, 013328 (2020)}.

\bibitem{KatoriNP} T. Akatsuka, M. Takamoto, H. Katori, Optical lattice clocks with non-interacting bosons and fermions
\href{https://doi.org/10.1038/nphys1108}{Nature Physics {\bf 4}, 954 (2008)}.

\bibitem{JYeScience2017} S. L. Campbell, R. B. Hutson, G. E. Marti, A. Goban, N. Darkwah Oppong, R. L. McNally, L. Sonderhouse, J. M. Robinson, W. Zhang, B. J. Bloom, and J. YeA, Fermi-degenerate three-dimensional optical lattice clock
\href{https://doi.org/10.1126/science.aam5538}{Vol 358, Issue 6359 (2017)}.

\bibitem{MultiManuel1} M. Gessner, L. Pezz{\'e}, and A. Smerzi, Sensitivity Bounds for Multiparameter Quantum Metrology,
\href{10.1103/PhysRevLett.121.130503}{Phys. Rev. Lett. {\bf 121}, 130503 (2018)}.

\bibitem{MultiManuel2} M. Gessner,  A. Smerzi and L. Pezz{\'e}, Multiparameter squeezing for optimal quantum enhancements in sensor networks, \href{https://doi.org/10.1038/s41467-020-17471-3}{Nat. Comm.  {\bf 11}, 3817 (2020)}.

\bibitem{ZwergerBook} BCS-BEC Crossover and the Unitary Fermi gas, ed. Wilhelm Zwerger,
\href{https://doi.org/10.1007/978-3-642-21978-8}{Springer Lecture Notes in Physics, {\bf 836}, (2011)}.

\bibitem{PRAphase} H. Kurkjian, Y. Castin, and A. Sinatra, Phase operators and blurring time of a pair-condensed Fermi gas,
\href{https://doi.org/10.1103/PhysRevA.88.063623}{Phys. Rev. A {\bf 88}, 063623 (2013)}.

\bibitem{CRphase} H. Kurkjian, Y. Castin, and A. Sinatra, Brouillage thermique d'un gaz coh{\'e}rent de fermions,
\href{https://doi.org/10.1016/j.crhy.2016.02.005}{Comptes Rendus Physique {\bf 17}, 789 (2016)}.

\bibitem{Obethaler1} J. Estève, C. Gross, A. Weller, S. Giovanazzi and M. K. Oberthaler, Squeezing and entanglement in a Bose–Einstein condensate, \href{https://doi.org/10.1038/nature07332}{Nature {\bf 455}, 1216 (2008)}.

\bibitem{Berrada2013} T. Berrada, S. van Frank, R. Bücker, T. Schumm, J.-F. Schaff and J Schmiedmayer, Integrated Mach–Zehnder interferometer for Bose–Einstein condensates, \href{https://doi.org/10.1038/ncomms3077}{Nature Communications {\bf 4}, 2077 (2013)}.

\bibitem{Lopez2017} R. Lopes, C. Eigen, N. Navon, D. Cl{\'e}ment, R. P. Smith and Z. Hadzibabic, Quantum Depletion of a Homogeneous Bose-Einstein Condensate, \href{https://doi.org/10.1103/PhysRevLett.119.190404}{Phys. Rev. Lett. {\bf 119}, 190404 (2017)}.

\bibitem{macQsimal}
macQsimal H2020-EU project, Miniature Atomic vapor-Cells Quantum devices for Sensing and Metrology Application
\href{https://www.macqsimal.eu/}{https://www.macqsimal.eu/}

\bibitem{MitchellNC2020} J. Kong, R. Jim{\'e}nez-Mart{\'i}nez2, C. Troullinou, V. G. Lucivero,
G\'eza T\'oth and M. W. Mitchell, Measurement-induced, spatially-extended entanglement in a hot, strongly-interacting atomic system \href{https://doi.org/10.1038/s41467-020-15899-1}{Nat. Comm. {\bf 11}, 2415 (2020)}.

\bibitem{GentileNacher2017}
T.~R. Gentile, P.~J. Nacher, B.~Saam, T.~G. Walker, Optically polarized $^{3}\mathrm{He}$,
\href{https://link.aps.org/doi/10.1103/RevModPhys.89.045004}{Rev. Mod. Phys. \textbf{89}, 045004 (2017)}.

\bibitem{Dantan2005a}
A. Dantan, G. Reinaudi, A. Sinatra, F. Lalo{\"e}, E. Giacobino, and M. Pinard, Long-lived quantum memory with nuclear atomic spins \href{https://doi.org/10.1103/PhysRevLett.95.123002}{Phys. Rev. Lett. {\bf 95}, 123002 (2005)}.

\bibitem{Reinaudi2007a}
G. Reinaudi, A. Sinatra, A. Dantan, and M. Pinard, Squeezing and entangling nuclear spins in helium 3, \href{https://doi.org/10.1080/09500340600677005}{J. Mod. Opt.  {\bf 54}, 675 (2007)}.

\bibitem{stroboscopic}
G.~Vasilakis, H.~Shen, K.~Jensen, M.~Balabas, D.~Salart, B.~Chen, E.~Polzik, Generation of a squeezed state of an oscillator by stroboscopic back-action-evading measurement, \href{https://doi.org/10.1038/nphys3280}{Nat. Phys. \textbf{11}, 389 (2015)}.  

\bibitem{FirstenbergPolzik}
O. Katz, R. Shaham, E. S. Polzik, and O. Firstenberg,
Long-Lived Entanglement Generation of Nuclear Spins Using Coherent Light, \href{https://doi.org/10.1103/PhysRevLett.124.043602}{Phys. Rev. Lett. {\bf 124}, 043602 (2020)}.

\bibitem{SerafinPRL2021}
A. Serafin, M. Fadel, P. Treutlein, A. Sinatra, Nuclear spin squeezing in Helium-3 by continuous quantum nondemolition measurement, \href{https://doi.org/10.1103/PhysRevLett.127.013601}{Phys. Rev. Lett 127, 013601 (2021)}.

\bibitem{SerafinCR2021}
A. Serafin, Y. Castin, M. Fadel, P. Treutlein, A. Sinatra, Etude th{\'e}orique de la compression de spin nucl{\'e}aire par mesure quantique non destructive en continu, \href{https://doi.org/10.5802/crphys.71 }{Compte Rendus Physique {\bf 22}, 35 (2021)}.

\bibitem{KornackRomalis} 
T. W. Kornack and M. V. Romalis, Dynamics of Two Overlapping Spin Ensembles Interacting by Spin Exchange,
\href{https://doi.org/10.1103/PhysRevLett.89.253002}{Phys. Rev. Lett. {\bf 89}, 253002 (2002)}.

\bibitem{OferNatPhys}
R. Shaham, O. Katz, O. Firstenberg, Strong coupling of alkali spins to noble-gas spins with hour-long coherence time,
\href{https://arxiv.org/abs/2102.02797}{https://arxiv.org/abs/2102.02797}.

\bibitem{LaloeDupontLeduc}
J.~Dupont-Roc, M.~Leduc, F.~Lalo{\"e}, Contribution {\`a} l'{\'e}tude du pompage optique par {\'e}change de m{\'e}tastabilit{\'e} dans 3He. - Premi{\`e}re Partie, \href{https://hal.archives-ouvertes.fr/jpa-00208119}{Journal de Physique \textbf{34}, 961 (1973)}.

\bibitem{TsePRL2019} M. Tse \textit{et al.}, Quantum-Enhanced Advanced LIGO Detectors in the Era of Gravitational-Wave Astronomy, \href{https://doi.org/10.1103/PhysRevLett.123.231107}{Phys. Rev. Lett. \textbf{123}, 231107 (2019)}; F. Acernese \textit{et al.}, Increasing the Astrophysical Reach of the Advanced Virgo Detector via the Application of Squeezed Vacuum States of Light, \href{https://doi.org/10.1103/PhysRevLett.123.231108}{Phys. Rev. Lett. \textbf{123}, 231108 (2019)}.

\end{thebibliography}

\end{document}




\title{Stati squeezed di spin per la metrologia} 



\author{Alice Sinatra}
\email[]{alice.sinatra@lkb.ens.fr}
\homepage[]{http://www.phys.ens.fr/~sinatra/}
\affiliation{Laboratoire Kastler Brossel, ENS-Universit\'e PSL, CNRS, Universit\'e de la Sorbonne et Coll\`ege de France,
24 rue Lhomond, 75231 Paris, France}


\date{\today}

\begin{abstract}
Lo spin-squeezing è una ``tecnologia quantistica" ormai consolidata, ove delle correlazioni ben disegnate in un insieme di sistemi a due livelli riducono l'incertezza statistica degli esperimenti spettroscopici. Questo articolo passa in rassegna alcuni progressi importanti nel campo, 
{ ponendo l'accento sui contributi dell'autore, in particolare sui limiti fondamentali imposti dalla decoerenza. Basandosi sul materiale presentato nella prima parte, nell'ultima sezione vengono delineate nuove idee ed alcuni sviluppi promettenti.}      
\end{abstract}

\pacs{}

\maketitle 

\section{Introduzione e panoramica}
\subsection{Spin collettivo, spin-squeezing e guadagno quantistico}
Un sistema quantistico a due livelli d'energia $E_a$ et $E_b$, è descritto formalmente come uno spin $1/2$. A partire da $N$ tali sistemi identici, per esempio degli atomi, si può formare un operatore di spin collettivo $\vec{S}$ le cui tre componenti in prima quantificazione hanno la forma
\begin{equation}
\hat{S}_x+i\hat{S}_y={\small{ \sum_{i=1}^N}} | a\rangle \langle b |_i \:;\:
\hat{S}_z=\frac{\sum_{i=1}^N (|a\rangle \langle a|_i-|b\rangle \langle b|_i)}{2}
\end{equation}

In uno stato del sistema in cui ogni atomo è in una sovrapposizione dei due livelli, lo spin collettivo precede attorno all'asse $Oz$ alla frequenza $\omega_{ab}=(E_b-E_a)/\hbar$. Che uno voglia misurare la frequenza $\omega_{ab}$ imperturbata di una certa transizione atomica, come in un orologio atomico, oppure che uno si interessi al cambiamento di tale frequenza indotto da un campo esterno, per esempio un campo magnetico, tale precessione è alla base di molti sensori atomici. 
\begin{figure}
\includegraphics[width=0.45\textwidth]{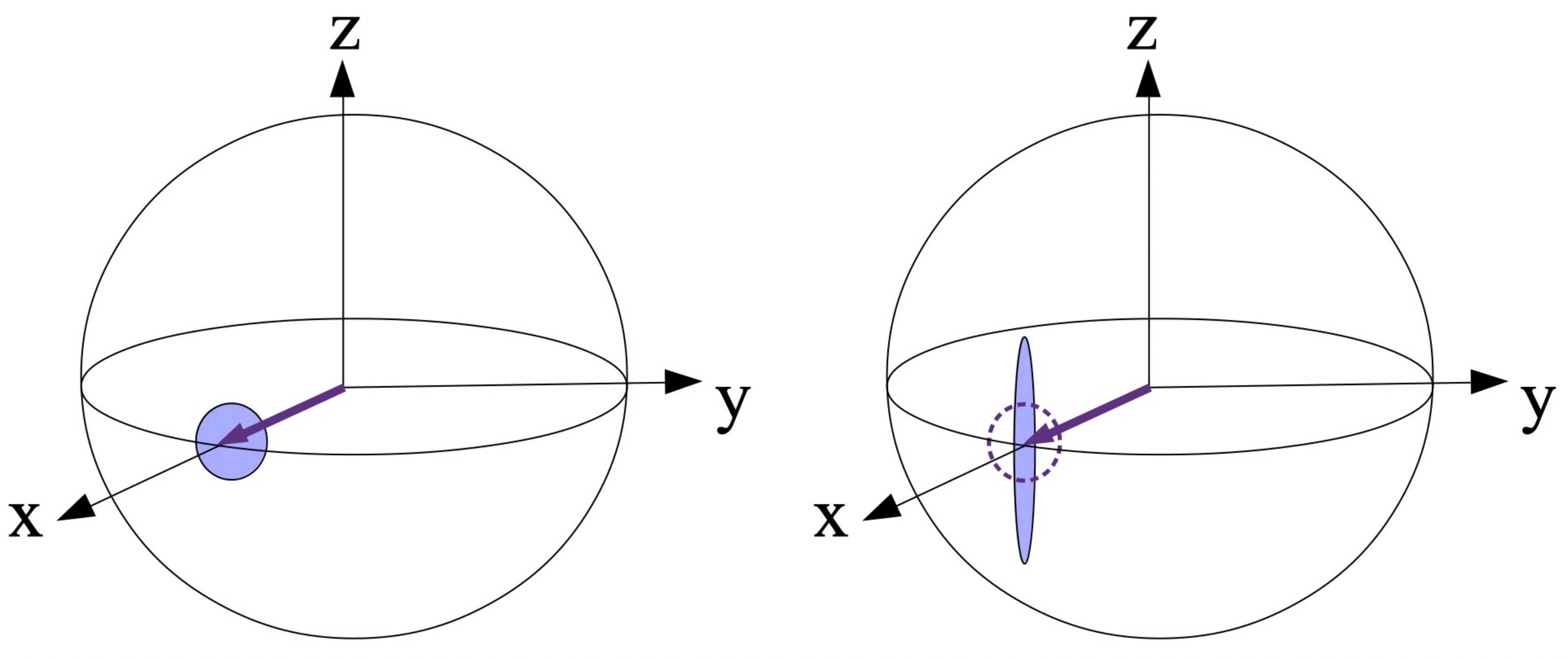}
\caption{Spin collettivo preparato nell'autostato di $S_x$ di valor proprio $N/2$ senza correlazione tra gli atomi (a sinistra) ed in uno stato spin-squeezed (a destra), rappresentato sulla sfera di Bloch di raggio $N/2$. La parte colorata in blu rappresenta schematicamente l'incertezza sulle componenti dello spin ortogonali allo spin medio, dovuta alla loro natura quantistica.
\label{fig:spin}}
\end{figure}
Tali sensori, anche dopo avere eliminato tutte le sorgenti di rumore tecnico, restano limitati dal rumore quantistico intrinseco allo stato in cui lo spin collettivo è stato preparato.  Se per esempio si prepara il sistema nell'autostato di $\hat{S}_x$ di valor proprio massimo $N/2$, stato non correlato prodotto tensoriale di $(|a\rangle+|b\rangle)/\sqrt{2}$ per ogni atomo, a sinistra nella figura \ref{fig:spin}, l'incertezza angolare sulla posizione dello spin collettivo è data da
\begin{equation}
(\Delta \phi)_{NC} = \frac{\Delta S_y}{\langle S_x \rangle} = \frac{\sqrt{N}}{N}=\frac{1}{\sqrt{N}} \,.
\end{equation}
Al contrario, in uno stato spin-squeezed \cite{KitagawaPRA1993}, a destra nella figura \ref{fig:spin}, l'incertezza su una delle componenti di spin ortogonali allo spin medio è ridotta grazie a delle correlazioni quantistiche tra gli atomi. Se l'incertezza ridotta è sulla variabile $S_y$ come in figura \ref{fig:spin}, la precisione $\Delta \phi$ sulla posizione angolare dello spin collettivo si trova aumentata. Si definisce allora un parametro di compressione di spin $\xi^2$, 
\begin{equation}
\xi^2 \equiv \frac{N\Delta S_\perp^2}{|\langle \vec{S} \rangle|^2}
\label{eq:xi2}
\end{equation}
sorta di rapporto tra rumore e segnale al quadrato in un'esperienza di spettroscopia, dove si vuole determinare la fase $\phi$, che si vuole più piccolo possibile e che permette di quantificare il guadagno metrologico di uno stato spin-squeezed rispetto ad uno stato non correlato. Si ha infatti\cite{WinelandPRA1992} 
\begin{equation}
\Delta \phi=\xi (\Delta \phi)_{\rm NC} =\frac{\xi}{\sqrt{N}}
\label{eq:phi}
\end{equation}
La definizione (\ref{eq:xi2}) può essere generalizzata per degli stati non gaussiani\cite{Gauss}, più complessi dello stato spin-squeezed rappresentato nella figura \ref{fig:spin}, per i quali l'informazione sulla fase di precessione $\phi$ deve essere estratta da una osservabile $\hat{X}$ che è in generale una funzione non lineare delle tre componenti dello spin collettivo\cite{GessnerPRL2019,Baamara2021,Baamara2022}, con un significato che rimane quello indicato dall'equazione (\ref{eq:phi}).
 
\subsection{Evoluzione non lineare e misura quantistica non distruttiva}
Per generare le correlazioni che portano da uno stato fattorizzato ad uno stato spin-squeezed, un metodo elegante consiste in una evoluzione unitaria con un hamiltoniana non lineare, per esempio\cite{KitagawaPRA1993} 
\begin{equation}
H_{\rm NL}=\hbar \chi \hat{S}_z^2
\label{eq:OAT}
\end{equation}
che rappresenta una non linearità di tipo Kerr per gli atomi. Essa può risultare dalle interazioni a corta portata tra atomi freddi in un condensato di Bose-Einstein\cite{SorensenNATURE2001,EPJD}, ma anche, in un insieme di atomi senza interazione, dalla loro interazione collettiva con un campo elettromagnetico in una cavità ottica \cite{VuleticPRA2010}. In entrambe i casi è possibile controllare il valore della costante $\chi$. Per i condensati, nel caso semplice in cui $a$ e $b$ giochino ruoli simmetrici e non interagiscano tra loro, $\chi=(d\mu/dN)/\hbar$ ove $\mu=\rho g$ è il potenziale chimico e $\rho$ la densità atomica per ciascuna delle due componenti. La costante di accoppiamento $g=4\pi\hbar^2a_s/m$  è proporzionale ad $a_s$, la lunghezza di scattering in onda $s$ tra gli atomi freddi.

Un altro metodo per creare uno stato spin-squeezed a partire da uno stato senza correlazioni consiste nell'effettuare una misura quantistica non distruttiva su una componente dello spin collettivo. In generale, per far ciò si accoppia il sistema $\mathcal{S}$ che si vuole misurare ad un secondo sistema quantistico $\mathcal{P}$ e, in seguito alla interazione tra i due sistemi, si effettua una misura distruttiva su $\mathcal{P}$.
Per misurare in modo non distruttivo la componente $\hat{S}_z$ dello spin collettivo, si può prendere come sistema secondario lo spin di Stokes $\hat{\vec{P}}$ { della luce} \cite{Stokes} 
ed utilizzare l'interazione di Faraday tra i due spin \cite{KuzmichEPL,Backaction}
\begin{equation}
H_{\rm Faraday}=\chi \hat{S}_z \hat{P}_z\,.
\label{eq:Fara}
\end{equation}
 
\subsection{Giro d'orizzonte sui risultati sperimentali e prospettive di applicazione}
Utilizzando i due metodi accennati nella  sezione precedente, gli stati spin-squeezed sono stati ottenuti in laboratorio e la dimostrazione del fatto che essi comportano un guadagno metrologico è stata fatta in diversi sistemi \cite{PezzeRMP2018}, che includono condensati di Bose-Einsetin gassosi\cite{GrossNATURE2010,RiedelNATURE2010}, atomi freddi interagenti con un campo elettromagnetico \cite{LerouxPRL2010,MitchellPRL2012,HostenNATURE2016} per i quali è stato misurato un guadagno metrologico $\xi^{-1}$ dell'ordine di $10$ con $5\times10^5$ atomi\cite{HostenNATURE2016}, e vapori atomici caldi \cite{BaoScience2020}. Questi risultati notevoli sono delle dimostrazioni di principio. Il passaggio alle applicazioni in metrologia dovrebbe aver luogo se il guadagno quantistico diventa facilmente accessibile dal punto di vista tecnologico o se in alcuni casi fosse indispensabile. Tra i vari sensori atomici, gli orologi che già nel 1999 avevano raggiunto ed osservato il limite quantistico nel caso di atomi indipendenti \cite{SantarelliPRL1999}, potrebbero essere i primi a beneficiare delle tecniche dello spin-squeezing. In generale, più il numero $N$ di atomi utilizzati è piccolo, pensiamo per esempio ad una evoluzione verso la miniaturizzazione dei sensori, più l'importanza relativa del rumore quantistico e la potenziale utilità degli stati spin-squeezed è grande. 

\section{Guadagno quantistico e decoerenza}
La metrologia quantistica si propone di migliorare la precisione delle misure riducendo la loro incertezza statistica ``fondamentale" data dalle fluttuazioni quantistiche. Ciò si ottiene preparando il sistema fisico di interesse in uno stato che presenta delle correlazioni quantistiche particolari. Affiché questo approccio abbia un senso da un punto di vista pratico, due condizioni ci paiono necessarie. 
La prima è che l'incertezza statistica sulla variabile, ottenuta da una misura con atomi indipendenti in un tempo impartito, sia dominata dalle fluttuazioni quantistiche e non da altre sorgenti di rumore classico. La seconda è che la decoerenza sia sufficientemente debole da non distruggere le correlazioni quantistiche che vogliamo mettere a profitto. Matematicamente, se si fissa il tempo $T$ della misura, la seconda condizione è automaticamente verificata nel limite $\gamma \to 0$ dove $\gamma$ è il tasso di decoerenza nel sistema.

\subsection{Leggi di scala del guadagno quantistico}
Se, come abbiamo detto, la decoerenza gioca un ruolo importante per l'utilizzo degli stati intricati, essa gioca ugualmente  un ruolo importante nella loro preparazione, determinando in generale un limite fondamentale al guadagno quantistico che si può ottenere.
Partendo da uno stato non correlato, autostato di $\hat{S}_x$ di valor proprio $N/2$ con $N\gg1$, l'evoluzione unitaria con l'hamiloniana non lineare (\ref{eq:OAT}) genera ad un tempo ottimale $\chi t \simeq N^{-2/3}$ uno stato spin-squeezed per il quale $\xi^2 \simeq N^{-2/3}$.
Questa legge di scala molto favorevole del guadagno quantistico per gli stati spin-squeezed diventa ancora più favorevole per stati non gaussiani ottenuti prolungando l'evoluzione unitaria per tempi più lunghi fino a raggiungere il limite massimo $\xi^2=N^{-1}$.\cite{ChalopinNCom2018,SchleierSmithPRL2016,VuleticArxiv2021} Tuttavia gli stati molto intricati che offrono un grande guadagno quantistico sono in generale più fragili rispetto alla decoerenza che cambia le leggi di scala del guadagno quantistico\cite{Kittens,Baamara2021,Baamara2022}. In alcuni casi si perde completamente la legge di scala e si ottiene un parametro di spin-squeezing che tende verso una costante $\xi^2_{\rm min}$ nel limite $N\to\infty$. Se questo indica che si ritrova una legge di scala classica per $\Delta \phi$ alla luce dell'equazione $\ref{eq:phi}$, il guadagno quantistico può rimanere molto grande se $\xi^2_{\rm min}\ll1$.

\subsection{Limiti fondamentali per lo spin-squeezing nei condensati di Bose-Einstein}

Con dei colleghi abbiamo studiato i limiti fondamentali della compressione di spin che si può ottenere in un gas condensato di Bose-Einstein formato da $N$ atomi a due livelli, solitamente due livelli iperfini nello stato fondamentale, utilizzando le interazioni di van der Waals a corto raggio tra atomi freddi. Se questo sistema fisico incarna bene in modello semplice (\ref{eq:OAT}) nella descrizione di campo medio\cite{SorensenNATURE2001,EPJD}, una sorgente intrinseca di decoerenza per la dinamica di spin-squeezing viene dall'interazione del condensato con le eccitazioni di bassa energia populate termicamente nel gas, la cui popolazione fluttua da una realizzazione all'altra dell'esperienza anche in condizioni identiche per i parametri sperimentali. Nel quadro di una descrizione microscopica multimodale, quella di Bogoliubov \cite{Bogoliubov}, abbiamo mostrato che al limite termodinamico $N\to \infty$ lo squeezing per un tempo di interazione ottimale $t_{\rm opt}$, che rimane finito, tende verso una costante $\xi^2_{\rm min}$ che costituisce quindi un limite per lo spin-squeezing\cite{SinatraPRL2011}
\begin{equation}
\xi^2(t_{\rm opt}) \stackrel{N\to\infty}{\to} \xi^2_{\rm min} \,.
\label{eq:xiopt}
\end{equation}
Per una temperatura più grande del potenziale chimico del gas, ma ben più piccola della temperatura critica di condensazione, il limite $\xi^2_{\rm min}$ è ben approssimato dalla frazione non condensata iniziale nel gas. Un punto importante è che mettendo dei valori realistici per i parametri, è in principio possibile ottenere di valori di $\xi_{\rm min}^2$ de l'ordine di $1/900$ corrispondente ad un guadagno di un fattore $30$ su $\Delta \phi$ secondo la formula (\ref{eq:phi}) rispetto all'utilizzo di atomi non correlati. 

Una seconda sorgente di decoerenza fondamentale che abbiamo considerato è data dalla perdita di particelle, conseguenza di un vuoto imperfetto per le perdite a un corpo, ed inerente alla natura metastabile dei gas ultrafreddi per le perdite a tre corpi. Nel quadro teorico di un sistema aperto descritto da una master equation\cite{LiYunPRL2008,SinatraFro2012} abbiamo mostrato che al limite di $N$ grande lo spin-squeezing ottimale è limitato da una costante, come in (\ref{eq:xiopt}), dove 
$\xi^2_{\rm min}$ è in questo caso proporzionale alla frazione di atomi perduti al tempo $t_{\rm opt}$. 
Una volta che lo spin-squeezing è stato preparato al tempo $t=0$, la sua evoluzione in assenza di interazioni ed in presenza di perdite a un corpo di tasso $\gamma$ è data da 
\begin{equation}
\xi^2(t)-1=\left[\xi^2(0)-1 \right]e^{-\gamma t} \,.
\label{eq:evloss}
\end{equation}
Dello spin-squeezing che dura più di un secondo ed è limitato dalle perdite secondo l'equazione (\ref{eq:evloss}) è stato visto in 
\cite{MylesArXiv}.

\section{Sviluppi promettenti per il futuro}
\begin{figure}
\includegraphics[width=0.45\textwidth]{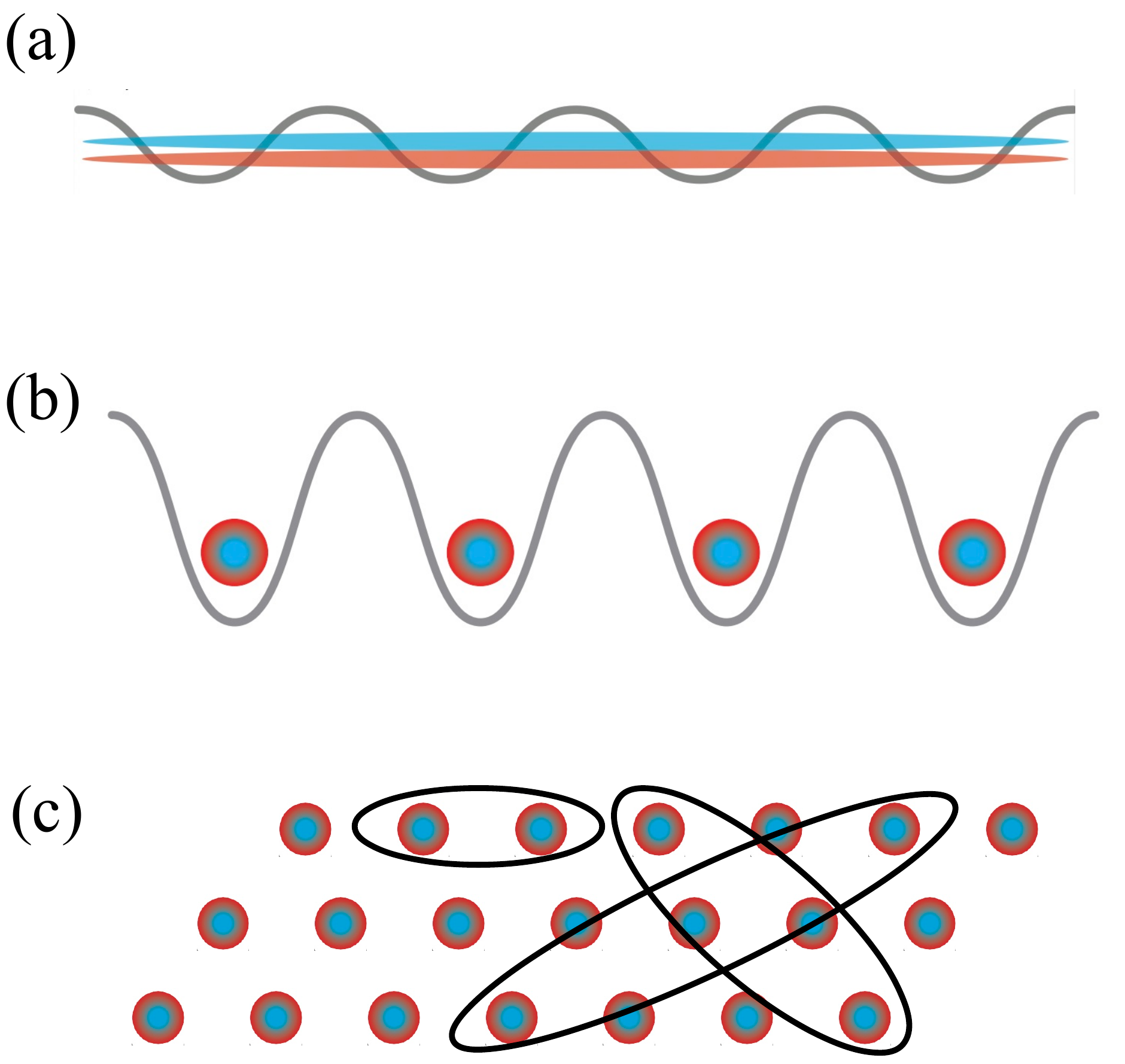}
\caption{(a) Si accende adiabaticamente un reticolo ottico in un condensato a due componenti in interazione. (b) Il sistema entra nella fase isolante di Mott con un atomo per sito at tempo ottimale di squeezing, congelando in tal modo le correlazioni quantistiche tra atomi. { (c) Nella configurazione squeezed-Mott, ogni atomo si trova in una sovrapposizione dei due stati interni $|a\rangle$ e $|b\rangle$ ed è ugualmente correlato a qualsiasi altro atomo. L'entanglement è quindi distribuito tra modi spazialmente separati e accessibili individualmente.} }
\label{fig:SqMott}
\end{figure}

{ La ricerca nel campo delle tecnologie quantistiche è abbondante, e diverse strade promettenti sono attualmente esplorate. Per gli orologi atomici, due esempi significativi sono la prima dimostrazione di spin-squeezing su una transizione ottica \cite{VuleticNat2020} e gli lo studi teorici di spin-squeezing con atomi nei reticoli ottici \cite{EPLSqMott,AMRlattice}. Lo scopo di questa sezione è di delineare alcuni sviluppi promettenti, a partire dal materiale che abbiamo presentato nella prima parte dell'articolo. Queste idee si inseriscono perfettamente nel quadro teorico descritto finora, in quanto si basano o sull'evoluzione non lineare (\ref{eq:OAT}) o sull'interazione quantistica di Faraday non-distruttiva (\ref{eq:Fara}). Tuttavia, a causa della differenza nel sistema fisico utilizzato, nelle scale temporali coinvolte o nel tipo di piattaforma quantistica che si può ottenere, la realizzazione di ciascuna di esse rappresenterebbe una svolta rispetto allo stato dell'arte e aprirebbe il campo a nuove applicazioni. 

{\it Stoccaggio dello spin-squeezed in una fase di Mott -}
Abbiamo già detto che le interazioni binarie di van der Waals a corto raggio in un condensato di Bose-Einstein dove ogni atomo è preparato in una sovrapposizione di due stati interni $(|a\rangle+|b\rangle)/\sqrt{2}$, generano l'hamiltoniana non lineare (\ref{eq:OAT}) che trasforma lo stato iniziale fattorizzato in uno stato di spin-squeezed. Fisicamente, poiché la forza di interazione dipende dallo stato interno degli atomi, la {\it fase di ogni atomo}, rappresentata dalla componente $\hat{S}_y$ dello spin collettivo, si correla grazie allo ``spostamento di campo medio" con la {\it differenza di popolazione} dell'insieme, rappresentata da $\hat{S}_z$. Per la misura di una fase o per l'interferometria atomica, lo squeezing, che è quindi generato in una sovrapposizione di $\hat{S}_y$ e $\hat{S}_z$, può essere trasferito alla componente di spin $\hat{S}_y$, come in figura \ref{fig:spin}, con una semplice rotazione. A questa descrizione fisica aggiungiamo ora un reticolo ottico che viene acceso adiabaticamente nel gas, portando progressivamente il sistema interagente dalla fase superfluida alla fase isolante di Mott \cite{Zoller1998}. Finché sussiste una componente superfluida delocalizzata, figura \ref{fig:SqMott} (a), la dinamica di spin-squeezing continua a creare correlazioni tra gli stati interni degli atomi. Quando il sistema entra nella fase Mott-insulator con un atomo per sito, figura \ref{fig:SqMott} (b), il condensato scompare, le interazioni di van der Waals sono soppresse poiché gli atomi sono separati, ma le correlazioni tra gli stati interni degli atomi che erano state create nello stato superfluido rimangono.}

Accendendo adiabaticamente un reticolo ottico in un condensato a due componenti dove le interazioni tra atomi  
incarnano la hamiltoniana (\ref{eq:OAT}), si può congelare uno stato spin-squeezed\cite{EPLSqMott}, o un stato non gaussiano più intricato ancora\cite{PRASqMott}, in una configurazione di isolante di Mott con un atomo per sito, figura \ref{fig:SqMott} (b), dove le interazioni di van der Waals sono soppresse poiché gli atomi sono separati, ed ogni atomo è uno spin $1/2$ che condivide con tutti gli altri delle correlazioni quantistiche. Oltre a l'interesse potenziale per gli orologi atomici\cite{KatoriNP,JYeScience2017}, una tale configurazione dove l'intricazione quantistica è condivisa tra modi spaziali individuamente accessibili offre delle prospettive interessanti per l'informazione quantistica e la metrologia quantistica multi-parametro\cite{MultiManuel1,MultiManuel2}. 

\begin{figure}
\includegraphics[width=0.5\textwidth,clip=]{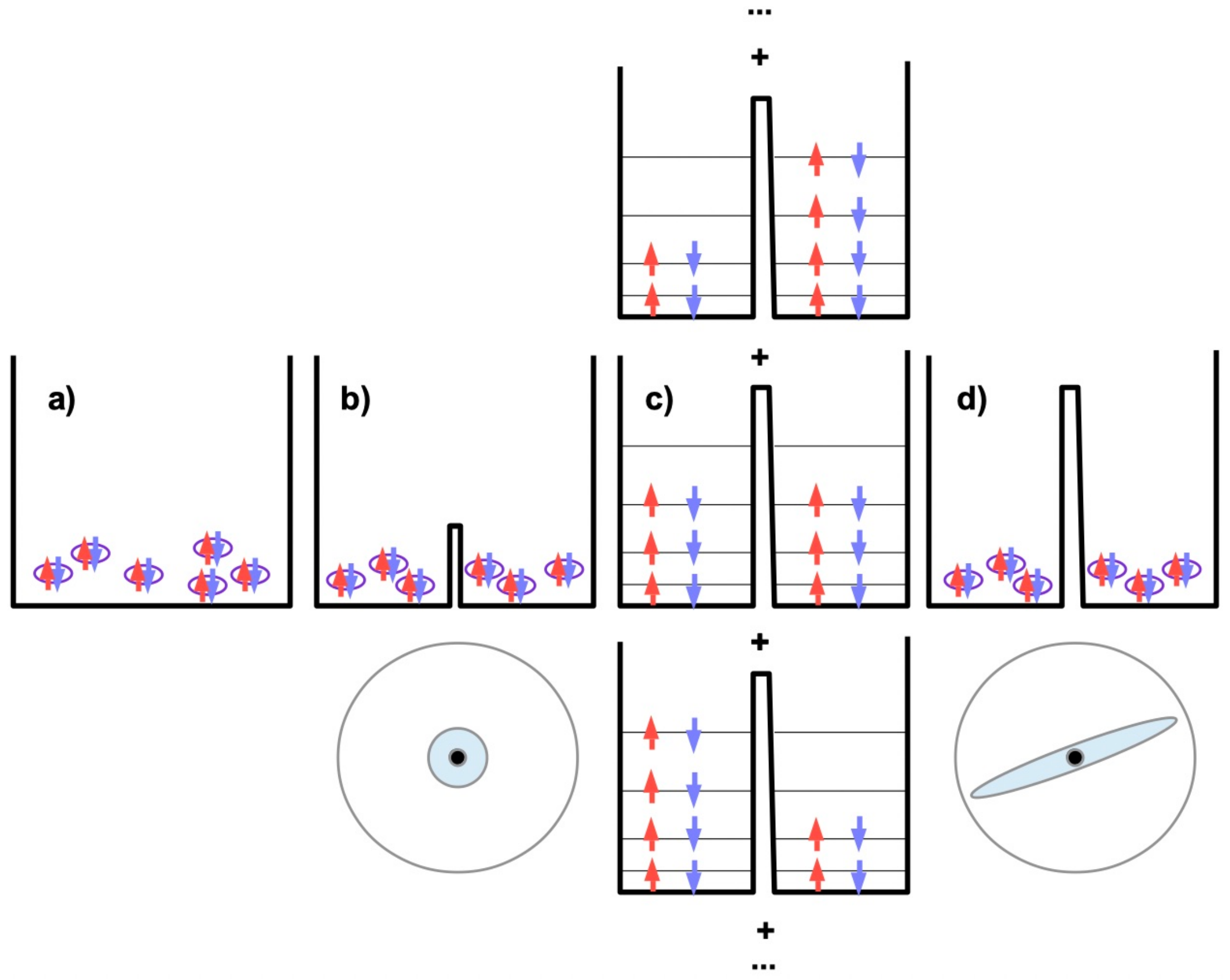}
\caption{Gedankenexperiment di spin-squeezing con un condensato di coppie di fermioni a temperatura nulla in una doppia buca. (a)-(b) un condensato di ``dimeri" di fermioni è preparato in uno stato coerente delocalizzato sulle due buche. (c) Dopo aver levato adiabaticamente la barriera separando le due buche, si porta la lunghezza di scattering al valore desiderato per esempio $a_s\to 0^-$ per la fase di spin-squeezing. Il questo caso lo squeezing si crea nel gas perfetto di fermioni grazie alla dipendenza da $N$ del potenziale chimico: $\epsilon_F \propto \rho^{2/3}$ a 3D. Dopo la fase di squeezing, si riporta la lunghezza di diffusione al valore
positivo iniziale corrispondente ad un condensato di dimeri, ottenendo per quest'ultimo uno stato spin-squeezed. Nelle parti inferiori delle figure (b) e (d) rappresentiamo le fluttuazioni dello spin collettivo, costruito a partire dai modi $a$ e $b$ del centro di massa dei dimeri corrispondenti allo stato fondamentale nella buca di destra e di sinistra, sulla sfera di Bloch.}
\label{fig:Fermi}
\end{figure}

{ \it Usare la non linearità di un gas degenere di fermioni -}
Delle prospettive interessanti sono anche offerte dai gas degeneri di fermioni in due stati di spin, nei quali si può controllare la forza delle interazioni senza indurre significative perdite di atomi, ed ottenere a piacere: un gas di dimeri formati da due atomi di spin opposto fortemente legati, un gas di coppie di Cooper debolmente legate, o un gas in interazione forte nel regime detto unitario \cite{ZwergerLibro}. 
Per darne un esempio, consideriamo il seguente {\it Gedankenexperiment} ispirato dai nostri studi sulla coerenza di fase di un condensato di coppie fermioni \cite{PRAphase,CRphase}. Un condensato di dimeri di fermioni di spin opposto è inizialmente preparato in una doppia buca simmetrica in uno stato non correlato, ove ogni è dimero simmetricamente delocalizzato sulle due buche (figura \ref{fig:Fermi}(a)-(b)). Alzando la barriera tra le due buche fino a separarle completamente, abbastanza lentamente per non eccitare i condensati, si prepara di fatto lo stato iniziale $\langle \hat{S}_x \rangle=N/2$ dove questa volta $a$ e $b$ si riferiscono alle due buche di potenziale, e si può mostrare che lo stato del sistema evolve secondo la hamiltoniana (\ref{eq:OAT}) ove, come per i condensati atomici $\chi=(d\mu/dN)/\hbar$, $\mu$ essendo il potenziale chimico del gas in ciascuna buca.
{ Si noti che qui $\hat{S}_z$ rappresenta la differenza tra il numero di coppie nelle due buche, $\hat{S}_x$ rappresenta la coerenza di fase macroscopica tra le due buche, e di nuovo, la quantità che viene {\it squizzata} dall'hamiltoniana (\ref{eq:OAT}) è una sovrapposizione di $\hat{S}_z$ e $\hat{S}_y$.}
Degli esperimenti simili con dei condensati di bosoni volti ad utilizzare gli stati spin-squeezed in un interferometro atomico sono stati realizzati \cite{Obethaler1,Berrada2013}. Nel caso del condensato di fermioni, come mostrato per esempio in \cite{PRAphase}, il valore più grande di $\chi$, quindi la dinamica di spin-squeezing più rapida, si ottiene nel limite $a_s\to 0^-$ dove il potenziale chimico in ciascuna buca tende verso l'energia di Fermi di un gas perfetto di fermioni $\mu \to \epsilon_F$. Questa situazione paradossale, ove lo spin-squeezing sarebbe generato in un mare di Fermi senza interazione, è rappresentata in figura \ref{fig:Fermi}(c).
Nel caso omogeneo a tre dimensioni è semplice paragonare le non linearità $\chi_{\rm Bose}=\frac{\rho g}{N}$ e $\chi_{\rm Fermi}=\frac{2}{3}\frac{\epsilon_F}{N}$ con $\epsilon_F\propto \rho^{2/3}$ e si vede che la non linearità è ben più grande per il gas perfetto di fermioni
\begin{equation}
\frac{\chi^{\rm Fermi}}{\chi^{\rm Bose}}\simeq 
\frac{\epsilon_F}{\rho g}\simeq 
\frac{1}{a_s\rho^{1/3}}\gg 1
\end{equation}
proporzionalmente all potenza $-\frac{2}{3}$ del parametro $\sqrt{\rho a_s^3}$, che tipicamente vale $10^{-3}$-$10^{-2}$ in un condensato di Bose-Einstein atomico \cite{Lopez2017}. Se queste considerazioni sono molto incoraggianti, uno studio a temperatura non nulla è necessario per trarre delle conclusioni sul migliore regime di interazioni da utilizzare con i fermioni ed calcolare li massimo spin-squeezing che si può ottenere con questo metodo.

{ \it Spin-squeezing nucleare a lunga durata di vita a temperatura ambiente -}
Nel campo dei sensori atomici, lo spin-squeezing e le tecnologie quantistiche hanno vocazione ad estendersi nei sistemi di gas o vapori in cella a temperatura ambiente\cite{MitchellNC2020,macQsimal}. {  Anche se il sistema fisico è diverso da quelli discussi finora, una volta presa la media sul movimento rapido degli atomi nella cella, è possibile descrivere i gradi di libertà interni in termini di spin collettivo e ricollegarsi alla descrizione teorica presentata nell'articolo.} Un sistema estremo per suo il livello di isolamento ed il tempo di coerenza che si misura in decine di ore, è lo spin collettivo puramente nucleare di un gas raro per esempio l'elio-3 nel suo stato fondamentale. Un primo studio teorico a mostrato che le collisioni di scambio tra gli atomi di elio nello stato fondamentale ed in uno stato metastabile accessibile otticamente, abitualmente utilizzate per polarizzare gli spin nucleari \cite{GentileNacher2017}, permettono anche di trasferire le correlazioni quantistiche e lo squeezing verso lo spin nucleare \cite{Dantan2005a,Reinaudi2007a} . Più recentemente è stato mostrato che degli schemi di spin-squeezing via una misura quantistica non distruttiva usando una hamiltoniana di Faraday di tipo (\ref{eq:Fara}), realizzati con successo per i gaz alcalini \cite{stroboscopic}, possono essere adattati allo squeezing dello spin nucleare \cite{FirstenbergPolzik,SerafinPRL2021,SerafinCR2021}, usando le collisioni di scambio di spin tra il gas raro e un alcalino \cite{KornackRomalis,OferNatPhys} oppure le collisioni di scambio di metastabilità in un gas puro di elio-3. \cite{LaloeDupontLeduc} In quest'ultimo caso, abbiamo mostrato che il massimo spin-squeezing $\xi^2$ ottenibile teoricamente\cite{SerafinCR2021} scala come $\sqrt{\gamma_0/\gamma_m}$ dove $\gamma_m$ è il tasso di collisoini di scambio di metastabilità per un atomo di elio metastabile, dell'ordine de $5\times10^6s^{-1}$ per una pressione di elio nello stato fondamentale di $2$ mbar, e $\gamma_0$ è il tasso di decohérenza dei metastabili, dominato dalla loro diffusione verso le pareti della cella, dell'ordine di $2\times10^4s^{-1}$ per una cella centimetrica senza gas tampone. Le numerose applicazioni spettroscopiche dello spin nucleare dell'elio, che vanno dalla magnetometria  ai test di fisica fondamentale \cite{GentileNacher2017}, potrebbero quindi in futuro beneficiare del controllo del rumore quantistico ed alimentare nuove idée ed applicazioni delle tecnologie quantistiche. 

{
\section{Conclusione}
Lo spin squeezing e la metrologia quantistica sono un campo molto attivo, costantemente alimentato da nuove idee, studi teorici approfonditi e sorprendenti progressi sperimentali. Il passaggio dalla scienza alla tecnologia dovrebbe avvenire quando il vantaggio quantistico supera la complessità degli esperimenti o, come abbiamo detto, se in alcuni casi fosse indispensabile. Un esempio interessante da questo punto di vista è quello degli stati squeezed della luce, ampiamente studiati negli anni '80 e ora utilizzati in un interferometro per rilevare le onde gravitazionali \cite{TsePRL2019}.}


%
%

%

\begin{acknowledgments}
Ringraziamo Yvan Castin per i suoi preziosi commenti sul manoscritto.
\end{acknowledgments}
